# SUBSTRATE SENSITIVITY OF THE ADHESION AND MATERIAL PROPERTIES OF RF-PECVD AMORPHOUS CARBON

SHASHI PAUL and F.J CLOUGH, The Emerging Technologies Research Centre, Department of Electrical & Electronic Engineering, De Montfort University, Leicester LE1 9BH, UK, pshashi@dmu.ac.uk.

## ABSTRACT

Hydrogenated amorphous carbon (a-C:H), deposited by the rf-plasma enhanced chemical vapour deposition (rf-PECVD) technique, is a promising material for large area electronic and interlayer dielectric applications. The structural and electronic properties of rf-PECVD a-C:H, deposited at room temperature from $CH_4$/He and $CH_4$/Ar gas mixtures, are shown to be sensitive to the substrate on which the thin film is deposited. The choice of substrate (c-Si or C7059 glass), and the existence and geometrical dimensions of any metallic pattern on the substrate surface, can result in significant spatial variations in the a-C:H adhesion and material properties. The observed effects are attributed to potential variations across the metal patterned substrates which influence the 'local' dc self-bias. This leads to spatial variations in the growth conditions and hence material properties. For electronic device and dielectric isolation applications this effect can result in significant variations in operating performance. The nature of the substrate and any overlying metallisation pattern are therefore important considerations.

## INTRODUCTION

Hydrogenated a-C films have received considerable attention over the last decade due to their unique diamond-like properties [1]. Low pressure, rf-PECVD is one of the most widely used techniques to deposit such films. By varying the deposition conditions, layers of a-C:H can be deposited at room temperature with properties ranging from semiconducting to insulating.

Rf-PECVD a-C:H has been applied to a wide variety of applications which utilise the material's electronic, rather than mechanical, properties. Such applications include metal-semiconductor-metal (MSM) [2] and thin film transistor (TFT) [3] switches for large area electronics and 'low k' dielectric isolation in VLSI and ULSI circuitry [4][5][6].

The interest in a-C:H for large area electronic applications arises from the ability to deposit semiconducting and insulating layers at room temperature. Such materials will therefore be fully compatible with the next generation of flat panel displays on plastic. Diamond-like carbon (DLC) and fluorinated DLC (F-DLC) have also attracted much attention for use in interlayer isolation. The delay time associated with an integrated circuit is a key measure of performance. 'Low k' dielectrics (i.e. materials with a dielectric constant much lower than that of $SiO_2$) are needed to minimise parasitic capacitances in ULSI circuitry. The combination of low dielectric constant, smoothness, mechanical strength and chemical inertness make a-C:H an ideal candidate for this application.

The critical growth parameter when depositing a layer of rf-PECVD a-C:H is the dc-bias which develops between the substrate and the plasma sheath [7]. The potential for the 'effective' dc-bias to be modified by the electrical properties of the substrate and the conductivity and topography of additional layers on the substrate surface is clear but, to date, no extensive investigation has been carried out. Such effects are likely to be particularly relevant when optimising layers of a-C:H for use in large area electronic and interlayer dielectric applications. The requirement to deposit layers of electronic grade a-C:H directly on top of narrow metallic lines (e.g. gate, source/drain or interconnect tracks) will make these applications very sensitive to such effects.

The present paper investigates the adhesion, morphology and electrical and structural properties of a-C:H films grown on a range of substrates (c-Si and C7059 glass) with and without pre-defined metallic patterns.

## EXPERIMENTAL

Thin films of a-C:H films were deposited at room temperature from $CH_4/Ar$ and $CH_4/He$ gas mixtures using a capacitively coupled rf-PECVD (13.56 MHz) system at a pressure of 100 mTorr. A nominal a-C:H film thickness of 90 nm was maintained for all samples. The plasma power was varied to develop dc self-bias voltages ranging from 70 V to 240 V. Test structures were prepared by thermally evaporating 20 nm thick films of Cr on to c-Si and C7059 glass substrates. Thin Cr layers were used to minimise step coverage induced stress effects in the deposited layers. Substrates (2 cm x 2 cm) with a continuous metal coating were used to investigate the properties of a-C:H films deposited on to large area metallic substrates. In addition, thin metallic strip and pad structures, with feature sizes up to several mm, were patterned on to selected substrates using shadow masks during the thermal evaporation stage. All substrates were subjected to an in situ Ar plasma pre-clean for 30 sec prior to the deposition of the a-C:H layer.

The current-voltage (I-V) characteristics of the a-C:H layers were measured using a purpose-built micro-tip system, connected to a pc-driven HP4140B pico-ammeter/voltage source. The micro-tip arrangement was installed within the chamber of a scanning electron microscope (SEM) and allowed the surface morphology and local electrical properties to be probed with high spatial precision [8]. Large area MSM structures were also produced for electrical characterisation by thermally evaporating 1 mm diameter metal dots on top of the a-C:H layer. Micro-Raman measurements were carried out using the 514.5 nm line of an Ar ion laser with a 2 µm spot size.

## RESULTS AND DISCUSSION

### Film Adhesion

A 90 nm thick a-C:H layer was deposited at a dc-bias of 210 V on to an un-coated c-Si substrate and a fully Cr coated c-Si substrate. The SEM image shown in Fig. 1(a) is for the fully Cr coated sample and reveals that the film is smooth and continuous with no visible pinholes. The a-C:H film deposited straight on to the c-Si substrate is of equally high quality. Hydrogenated a-C films were next grown, under identical conditions, on c-Si substrates coated with 20 nm thick Cr-strips of dimensions 1 mm x 20 mm. The poor adherence of the a-C:H layer to the Cr coated part of the substrate is clearly shown in Fig. 1(b). In contrast, the a-C:H film on the un-coated regions of the c-Si substrate is continuous and homogenous (top left in Fig. 1(b)).

Some reduction in film adhesion is predictable at the edge of the Cr pad, where there is a significant change in the surface topography and film stress will be enhanced. However, this does not explain the general observation of excellent film adherence to the c-Si (up to the step edge) and the poor adhesion on the Cr coated regions situated many microns from the step edge. It is quite clear from our investigations that under certain deposition conditions the a-C:H layer adheres preferentially to the c-Si rather than the metallic coated regions of the substrate. Such observations confirm that the substrate material and its surface topography are important considerations when optimising the a-C:H growth conditions for a particular application and device geometry.

To investigate further, we systematically studied the adherence of a-C:H across c-Si substrates with pre-defined Cr strips, at different values of the dc self-bias. The results show a well defined 'critical dc-bias' of 200V for a-C:H films deposited in our rf-PECVD system. Films deposited at dc biases less than 200 V show good adhesion to both the exposed c-Si surface and the Cr strips. A typical SEM image of a film deposited below the 200V is shown in the Fig. 2. Films deposited above 200V stick to the c-Si regions but rapidly peel away from the Cr tracks.

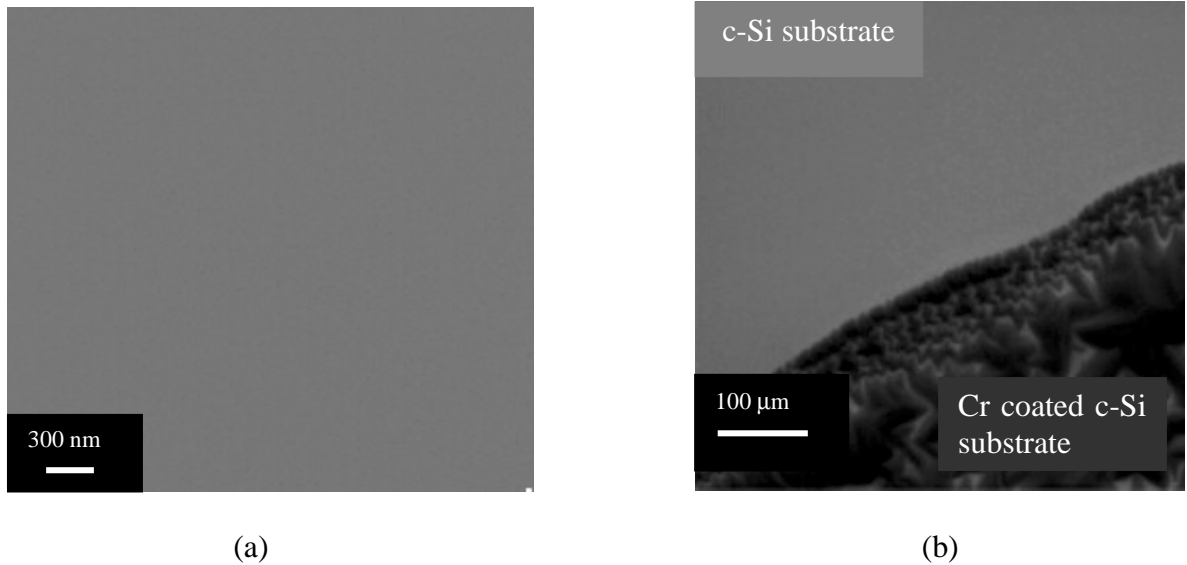

(a)          (b)

Fig.1 (a) SEM micrograph of an a-C:H film deposited on to a continuous Cr-coated substrate and (b) SEM micrograph of an a-C:H film deposited on to a c-Si substrate with a pre-defined Cr Strip.

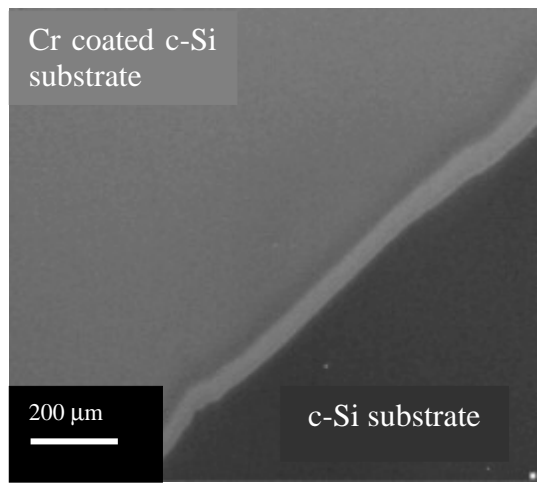

Fig. 2. SEM image of an a-C:H layer covering adjacent c-Si and Cr coated regions.

Adhesion is closely related to the film stress and the dependence of film stress on the dc bias is well documented for rf-PECVD a-C:H [7]. Within the dc bias range which results in 'hard' a-C:H films, an increase in the dc bias voltage results in a corresponding increase in the film stress. It is suggested that the observed reduction in film adhesion on the metallic tracks can be attributed to field enhancement in the region around the metallic patterns arising from their high conductivity (relative to the surrounding c-Si substrate). Field enhancement results in a 'local' dc bias which can exceed the 'average' dc bias set by the process controller. The corresponding increase in the film stress is then sufficient to overcome the cohesive force between the metal strip and the a-C:H film and the film peels off. This rationale is supported by the observation of poor film adhesion on fully Cr coated c-Si substrates at higher 'average' applied dc biases. Work is currently ongoing within our laboratory to develop a detailed, quantitative analysis of the change in the electric field distribution associated with a surface metallic pattern.

**<u>Material Properties</u>**

We have recently developed a technique by which the electrical characteristics of a wide range of materials can be measured, with great spatial precision, using a combination of SEM and a sub-micron tip [8]. This technique was used to investigate the electrical properties of continuous a-C:H layers (i.e. with good adhesion) at a variety of locations in the vicinity of the Cr tracks. Typical I-V characteristics for an a-C:H film deposited at a dc bias of 150V on top of the Cr strip are shown in Fig. 3. Fig. 3 (a) shows an I-V characteristic of a 'micro-tip' MSM structure formed in the middle of the strip. The I-V characteristic is symmetrical for negative and positive applied voltages and is indicative of a bulk-limited Poole-Frenkel conduction mechanism [2][9]. Fig. 3 (b) shows the typically asymmetric I-V characteristic of a 'micro-tip' MSM formed near the edge of the Cr Strip.

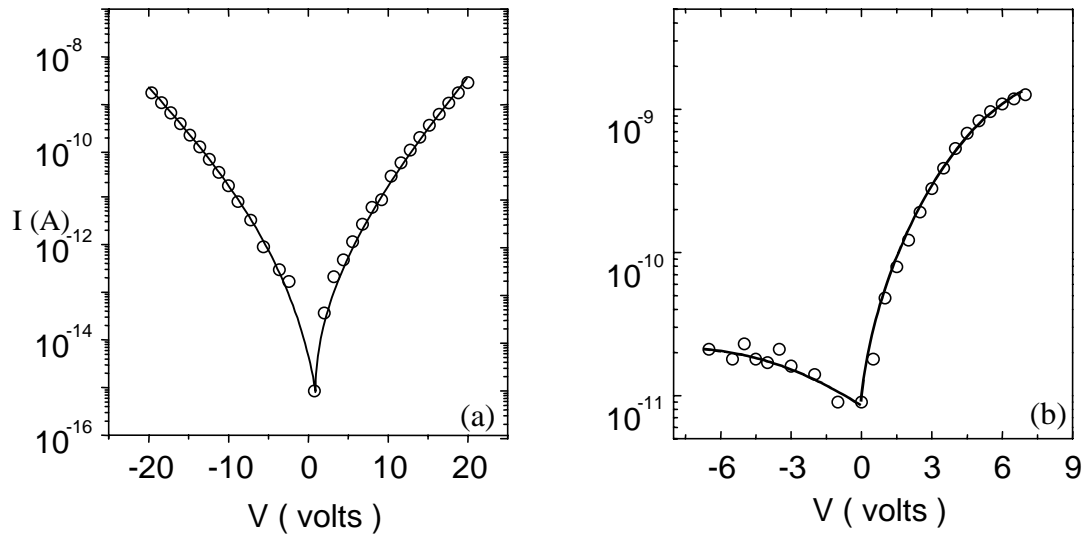

<u>Fig. 3</u>. (a) A typical symmetrical I-V characteristic for a 'micro-tip' MSM structure in the middle of a Cr strip (b) A typical asymmetrical I-V characteristics from a 'micro-tip' MSM structure at the edge of a Cr strip.

We have previously reported similar inhomogeneities in the I-V characteristics of 'micro-tip' M/a-C:H/M structures formed at different locations on a continuous metal coated c-Si substrate [9]. The observation of symmetrical I-V behaviour was attributed to $sp^2$ rich regions in the a-C:H film. The observation of asymmetric I-V behaviour was attributed to Schottky barrier formation at $sp^3$ rich regions of the film. Of key significance in the present work is the high incidence of asymmetric I-V characteristics near the strip edge (high $sp^3$ content) and symmetrical I-V characteristic in the centre of the strip (high $sp^2$ content). This observation is a clear indication that the material properties of the a-C:H layer vary from the middle to the edge of the Cr strip. It is again proposed that the observed spatial variation in the material properties is associated with a higher 'local' dc bias in the vicinity of the strip edge.

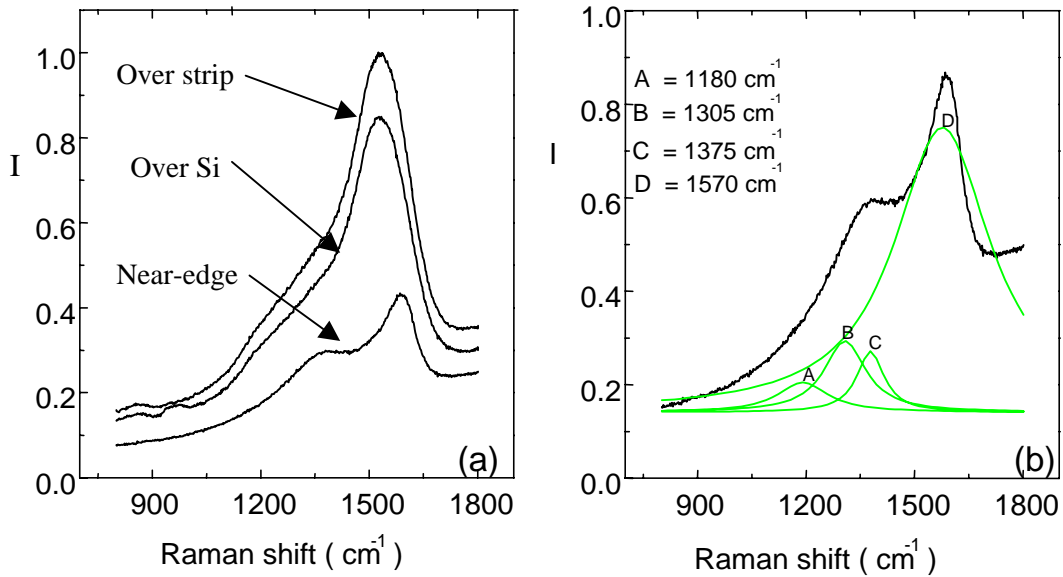

Fig. 4. (a) Typical Raman spectra of a-C:H at three different locations (b) de-convoluted Raman spectrum collected from an a-C:H region near the edge of a metal strip.

Micro-Raman spectroscopy has been carried out to confirm the structural variation in the a-C:H film across the metal tracks. Typical Raman spectra, extracted from the same a-C:H layer near the edge of a metal-strip, in the centre of a metal strip and over the c-Si substrate, are shown in Fig. 4 (a). The Raman spectrum at the edge of the metal strip is clearly and reproducibly different. Structural information can be extracted by de-convoluting the spectra into a series of Lorentzian peaks. De-convolution of the Raman spectrum obtained from the track edge results in four individual peaks centred at approximately 1180 cm$^{-1}$, 1305 cm$^{-1}$, 1375 cm$^{-1}$ and 1570 cm$^{-1}$. The peaks at 1375 cm$^{-1}$ (D peak) and 1570 cm$^{-1}$ (G peak) can be attributed to $sp^2$-type bonding configurations [10]. The slight shift in these peaks may be due to the high stress in the film, near the edge of the metal track. The peaks at 1180 cm$^{-1}$ and 1305 cm$^{-1}$ are typically attributed to nano-crystalline diamond or, $sp^3$ rich carbon [10]. De-convolution of the a-C:H Raman spectra obtained from the centre of the metal strip and the un-coated c-Si substrate results in D and G peaks only.

This observation is consistent with the preceding electrical measurements and is further strong evidence that the a-C:H material properties are sensitive to the underlying substrate.

## CONCLUSIONS

The adhesion and material properties of thin films of rf-PECVD a-C:H film can be significantly affected by the substrate material on which the film is deposited and the substrate topography. For applications in large area electronics and interlayer isolation these issues will require serious consideration. We have investigated the material variations associated with metallic films of a relatively small aspect ratio (20 nm thick and 1 mm wide). In real applications the metal tracks may be considerably thicker (up to a few microns) and narrower (down to a few microns). This will result in more extreme variations in the 'local' dc bias and hence the material properties. A better understanding of the substrate dependence of the 'local' dc bias and the microscopic properties of a-C:H is therefore important if these materials are to be reliably integrated into large area electronic and 'low-k' dielectric applications.

## ACKNOWLEDGEMENTS

The authors would like to thank Renishaw plc. for the use of their Raman micro-probe system and in particular Mike Cranbourne for his help in collecting and interpreting the Raman data.